# Koopman Spectral Analysis of Intermittent Dynamics in Complex Systems: A Case Study in Pathophysiological Processes of Obstructive Sleep Apnea

Phat K. Huynh[a], Arveity R. Setty[b,c], and Trung Q. Le[a,d]*

[a] *Department of Industrial and Manufacturing Engineering, North Dakota State University at Fargo, North Dakota, USA;* [b]*Sanford hospital in Fargo, North Dakota, USA;* [b]*University of North Dakota, North Dakota, USA;* [d]*Department of Biomedical Engineering, North Dakota State University at Fargo, North Dakota, USA*

*\*Corresponding author: Trung. Q. Le, PhD, is with North Dakota State University, ND 58102 (e-mail: trung.q.le@ndsu.edu)*

**Phat K. Huynh** is currently a PhD student of the Industrial and Manufacturing Engineering Department at North Dakota State University. He earned a bachelor's degree in biomedical engineering, International University - VNU-HCM in 2018, Viet Nam. His work mainly focuses on wearable devices, nonlinear dynamical systems, probabilistic statistical models, machine learning, and predictive analytics in healthcare.

**Arveity R. Setty** is a Sleep Medicine Specialist in Fargo, ND and has over 9 years of experience in his field. He graduated from Michigan State University, College of Human Medicine Medical School in 2012. He is affiliated with Sanford Health. He serves as a Clinical Associate Professor at the University of North Dakota School of Medicine & Health Services.

**Trung Q. Le** is an Assistant Professor of the Industrial and Manufacturing Engineering Department at North Dakota State University. He was an Assistant Professor and Associate Department Head for Research Affairs in Biomedical Engineering Department at International University - VNU-HCM, Vietnam. His research focuses in 3 main directions: 1) Data-driven and Sensor-based Modeling, 2) Medical Device Manufacturing and Bio-signal Processing, and 3) Predictive Analytics for Personalized Healthcare.

# Koopman Spectral Analysis of Intermittent Dynamics in Complex Systems: A Case Study in Pathophysiological Processes of Obstructive Sleep Apnea

**ABSTRACT**: Complex systems, such as pathophysiological processes, commonly exhibit chaotic, nonlinear, and intermittent phenomena. Koopman operator theory and Hankel alternative view of Koopman (HAVOK) model have been widely used to decompose the chaos of the complex system dynamics into an intermittent forced linear system. Although the statistics of the intermittent forcing have been proposed to characterize intermittencies in the HAVOK model, they were not adequate to attribute for the mode switching of nonlinear dynamics and the fat-tailed non-Gaussian distribution originated from high-frequency bursts and rarely-observed intermittent forcing. The paper proposed a new intermittency dynamics analysis approach to characterize the intermittent phases, chaotic bursts, and local spectral-temporal properties of various intermittent dynamics modes using spectral decomposition and wavelet analysis. To validate our methods, the intermittency behavior of apneic events in obstructive sleep apnea disorder was selected as the case, in which heart rate variability (HRV) features were extracted. Next, we constructed the Hankel matrix from the HRV features and obtained the last eigen time-delay coordinate by singular value decomposition of the Hankel matrix, which was modeled as an intermittent forcing input. The statistics of the forcing in OSA demonstrated the fat-tailed distribution of the intermittent forcing, which correspond to the intermittency of the underlying OSA pathophysiological process. The pooled means and standard deviations of the burst duration and the inter-burst duration across OSA patients were also calculated to be minutes and minutes. Scalogram amplitude and spectral decomposition of the wavelet transform exhibited various predominant frequencies and dynamics modes associated with apneic events.

## 1. Introduction

Modeling complex systems dynamics, which exhibit nonlinearity, chaos, intermittency, transient, and uncertainty (Guckenheimer & Holmes, 2013; Majda & Lee, 2014; Ye et al., 2015), has become one of the grand challenges in engineering and science (Brunton

& Kutz, 2019). Among these behaviors, intermittency is one of the most common phenomena observed in many complex processes including human pathophysiology processes (Shelhamer, 2007). The intermittent dynamics have been reported in neuromuscular events in tremor (Wang, Aziz, Stein, & Liu, 2005), beta-oscillations synchronization events in Parkinson's disease (Park, Worth, & Rubchinsky, 2010), and obstruction of the posterior nasopharynx in obstructive sleep apnea (Donnelly, Shott, LaRose, Chini, & Amin, 2004; Le & Bukkapatnam, 2016). Although the intermittency behaviors have been widely observed, the modeling and analysis of intermittency has been not well-developed in nonlinear complex systems especially with high-dimensional and nonlinear coupled physiological dynamic systems.

Intermittency is defined as the erratic alternations of the systems between periodic (regular, laminar) dynamics and chaotic (irregular, turbulent) behaviors characterized commonly by short bursts in the signal (Kantz & Schreiber, 2004). Intermittency can also exist in the other form of chaotic dynamics called crisis-induced intermittency (Grebogi, Ott, Romeiras, & Yorke, 1987). To clarify the origins of the intermittency phenomena, two groups of explanations have been proposed. First, intermittency may originate from Hamiltonian chaos (Zaslavsky & Zaslavskij, 2005) and hydrodynamical systems (Bessaih, Hausenblas, & Razafimandimby, 2015). Second, intermittent behaviors can also arise from small control parameter fluctuations around critical values (Contoyiannis, Diakonos, & Malakis, 2002; Hramov, Koronovskii, & Kurovskaya, 2014). To prevent the confusion between non-intermittent dynamics mode dynamics due to non-stationarity and the dynamical intermittency for fixed parameters of dynamical systems, the statistics of intermittent phases and chaotic bursts need to be studied thoroughly (Kantz & Schreiber, 2004). However, there are increasingly many complex systems for which we have abundant measurement data from sensors but do not have access to the underlying parameterized governing equations. As a result, the study of the intermittency dynamics in such systems are considerably challenging since the bifurcation analysis from the governing equations is the key approach to study the system behaviors as the parameters are varied. Hence, a data-driven method that can model the intermittent dynamics for high-dimensional complex systems effectively and interpretably is necessitated.

Cutting-edge algorithms in machine learning and big data have revolutionarily shifted from the theory-based analysis of dynamical systems towards data-driven-based analysis over the last decade, (Ashwin, Coombes, & Nicks, 2016; Brunton & Noack, 2015; Majda & Lee, 2014; Sapsis & Majda, 2013)). A pioneering work that contributes to the intermittent nonlinear chaotic dynamics analysis is Takens' delay embedding theorem (Takens, 1981), where the delay embeddings are used to reconstruct the state space and characterize intermittent nonlinear chaotic systems (Abarbanel, Brown, Sidorowich, & Tsimring, 1993; Sugihara et al., 2012; Ye et al., 2015). In the later years, the Takens' embedding theorem has also been applied in the eigensystem realization algorithm (ERA) for linear system identification (Juang & Pappa, 1985), the singular spectrum analysis (SSA) in climate science (Wei, 2006), and the nonlinear Laplacian spectrum analysis (Giannakis & Majda, 2012). However, these methods are still limited in modeling and interpreting the intermittent and nonlinear behaviors in complex systems. To overcome those limitations, Brunton et al. have developed a data-driven Hankel alternative view of Koopman (HAVOK) model (Brunton, Brunton, Proctor, Kaiser, & Kutz, 2017), which decomposes chaos into a forced linear system using regression models (Brunton, Proctor, & Kutz, 2016) and the modern Koopman operator theory (Giannakis, 2019) to decomposes chaos into a forced linear system and identify invariant subspaces for nonlinear systems with multiple attractors and periodic orbits. The method uses the linear models on eigen time-delay coordinates obtained from the Hankel matrix, which represent the dynamics in the attraction basins of the fixed points or periodic orbits using appropriate measurement functions (Lan & Mezić, 2013). This model is well-suited to model the nonlinear dynamics that are chaotic and intermittent, in which the trajectories of the system dynamics evolve to densely fill the attractor. The HAVOK model has been applied to decompose chaos of nonlinear systems including analytical systems (e.g., Lorenz and Rössler) with deterministic governing equations, stochastic magnetic field reversal, and real-world systems such as cardiovascular, neurological, and epidemiological systems (Brunton et al., 2017). Although the statistics of the intermittent forcing have been proposed to characterize intermittencies in these models, they were not adequate to characterize the mode switching of nonlinear dynamics and the fat-tailed non-gaussian distribution originated from high-frequency bursts and rarely-observed intermittent switching events (Kantz & Schreiber, 2004; Majda & Harlim, 2012; Majda & Lee, 2014). Therefore, the intermittent forcing interpretation in the HAVOK model has been not fully elucidated, and the connection between the complex nonlinear dynamics and the bursting of intermittent forcing has not been studied.

Therefore, our paper proposed an intermittency analysis method that systematically decomposes and analyzes the intermittent forcing obtained from the HAVOK model. Particularly, in addition to the distribution of the intermittent forcing, we proposed additional statistics such as burst starting-ending time, burst duration, and inter-burst duration quantifiers to better characterize the intermittent phases and chaotic bursts. Moreover, the temporal spectra of the intermittent forcing signal estimated from Fourier transform (DFFT) (Oppenheim, Wornell, Isabelle, & Cuomo, 1992) was used to characterize the changes in the spectral properties of the intermittent forcing (e.g., predominant frequency bands) and potentially explain the switching between different intermittency modes. The adaptive continuous-time wavelet analysis (Lilly, 2017) with different types of mother wavelets chosen to match the morphological features of the bursts was performed to extract local spectral and temporal information simultaneously. To validate our proposed methods, a sleep disorder, namely obstructive sleep apnea (OSA), was selected as the case study. The main contributions of our paper are (1) a systematic intermittency analysis method to characterize the intermittent phases, the chaotic bursts, and the spectral-temporal properties of different intermittent dynamics modes, (2) an attempt to characterizing chaos in pathophysiological processes as an intermittently forced linear and thoroughly analyze the intermittency components. The understanding of biological system's intermittency dynamics enhances the development of interpretable and robust domain-knowledge data-driven methods for complex systems that exhibit intermittent and chaotic behaviors. The estimation and prediction of the future system behaviors such as paroxysmal events can be carried out easily using the dynamical system obtained by our methods with data-tailored system coordinates and fine-tuned parameters. The remainder of the paper is organized as follows: Section 2 introduces about the background of dynamical systems, Koopman operator theory, Taken's delay embedding theory, and the HAVOK model; Section 3 elaborates the intermittent forcing analysis methods; the results of the intermittency analysis for the obstructive sleep apnea (OSA) case study are presented in Section 4; the discussion and conclusion are provided in Sections 5 and 6.

## 2. Background

Throughout this paper, we will consider continuous-time dynamical systems with state vector $x(t) \in \mathbb{R}^n$ in the form of:

$$\frac{d}{dt}x(t) = f(x(t)) \tag{1}$$

where $f(\cdot)$ denotes the flow map operator. Correspondingly, the discrete-time dynamical system is defined as:

$$x_{k+1} = F(x_k) \tag{2}$$

where $x_k = x(k\Delta t)$ is the sample of the system trajectory from Eq. (1). The discrete-time propagator $F$ is given by:

$$F(x_k) = x_k + \int_{k\Delta t}^{(k+1)\Delta t} f(x(s))ds \tag{3}$$

### 2.1. Koopman operator theory

The Koopman operator theory (Koopman, 1931) constructs a linear representation of strongly nonlinear systems by augmenting the $n$-dimensional state vector to an infinite-dimensional state vector. This linear representation is a global linearization of dynamical systems which is more advantageous than the local linearization and valid only near fixed points or periodic orbits (Wanner, 1995). The Koopman operator $\mathcal{K}$ is defined as an infinite-dimensional linear operator that acts on measurement functions (i.e., Koopman observables) $g: M \mapsto \mathbb{R}$ of the state $x$ where $M$ is the manifold. The Koopman operator is defined as:

$$\mathcal{K}g \triangleq g \circ F \Longrightarrow \mathcal{K}g(x_k) = g(x_{k+1}) \tag{4}$$

In Koopman operator theory, a linear subspace spanned by the measurement functions $\{g_1, g_2, \dots, g_p\}$ is considered so that for any measurement $g$ in this space, we have:

$$g = \alpha_1 g_1 + \alpha_2 g_2 + \cdots + \alpha_p g_p \tag{5}$$

and $g$ remains in the Koopman invariant subspace after being acted on by $\mathcal{K}$:

$$\mathcal{K}g = \beta_1 g_1 + \beta_2 g_2 + \cdots + \beta_p g_p \tag{6}$$

where $\alpha_k, \beta_k \in \mathbb{R}, \forall k = 1, \dots, p$ are the coefficients for the linear subspace span of the observables $g_k$ before and after applying $\mathcal{K}$. Here, $\mathcal{K}$ is restricted to $p$-dimensional measurement subspace $\mathbb{R}^p$ and represented by a $p \times p$ matrix $K$. If $K$ exists, we can obtain an exact linear system that advances the Koopman observables restricted to the subspace as follows:

$$y_{k+1} = Ky_k, \qquad y_k = [g_1(x_k), g_2(x_k), \dots, g_p(x_k)]^T \tag{7}$$

where $y_k$ is a vector of measurements in the Koopman invariant subspace as illustrated in Fig. 1.

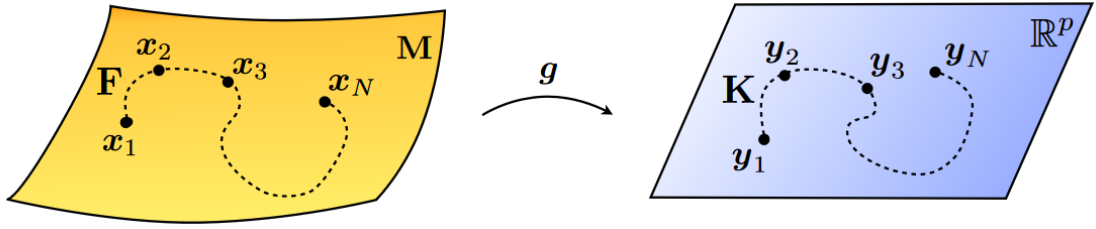

**Figure 1**. Schematic illustrating the capability of the Koopman operator $K$ to globally linearize a nonlinear system using an appropriate selection of observable measurement functions $g$ (Brunton et al., 2017).

In practical applications, there have been many attempts to obtain a finite-dimensional approximation of the Koopman operator $\mathcal{K}$, but they have had limited success. Dynamic mode decomposition (DMD) (Tu, 2013) was applied to obtain a best-fit linear matrix that advances spatial measurements from one time slice to the next; however, this best-fit linear matrix appears not to be a good approximation for the Koopman operator and yields a large error in many nonlinear systems. To overcome that, the DMD method was extended to include nonlinear measurements (Williams, Kevrekidis, & Rowley, 2015), but there is no existing theory that guarantees the extended model is closed under the Koopman operator (i.e., the observables advanced by $\mathcal{K}$ are restricted to the measurement subspace). Therefore, a better finite-dimensional approximation of $\mathcal{K}$ –which relies on a Koopman invariant measurement subspace was introduced (Brunton, Brunton, Proctor, & Kutz, 2016). Instead of using DMD or extended DMD, this model is obtained by using intrinsic data-driven measurement coordinates derived from the time-history of the system measurements, namely eigen time-delay coordinates. We will further discuss this model in Section 2.3.

## 2.2. Takens' embedding theorem for state space reconstruction

The Takens' embedding theorem (Takens, 1981) states we can enrich a measurement, $z(t)$, with time-shifted copies, $z(t - \tau)$, known as delay coordinates so that the attractor of a dynamical system is diffeomorphic to the original attractor under certain conditions. However, in real-life complex systems, we always observe discrete-time univariate time series $\{z_n\}_{n=1}^{N}$ of length $N$, the system dynamics can be obtained through an appropriate state space reconstruction from the time series. According to the theorem, a trajectory is formed by points $\hat{x}_j$ using the delay map as follows:

$$\hat{x}_j = \left(h(x_j), h(x_{j+\tau}), \ldots, h(x_{j+(m-1)\tau})\right) = (z_j, z_{j+\tau}, \ldots, z_{j+(m-1)\tau}), \quad (8)$$
$$j = 1, \ldots, N - m + 1$$

where $\tau$ is the time lag, $m$ is the embedding dimension, and $h: M \mapsto \mathbb{R}$ is the measurement function.

### 2.3. Modeling chaos as an intermittently forced linear system: Hankel alternative view of Koopman (HAVOK)

#### 2.3.1. State space reconstruction by the decomposition of the Hankel matrix

Chaotic dynamics are remarkably well-matched with the proposed data-driven HAVOK analysis (Brunton et al., 2017) because the trajectories in a chaotic system will evolve to densely fill an attractor. Therefore, the longer time-history of the system we obtain, the more information it provides to characterize the chaotic dynamics of the system. The first step and second step in the HAVOK analysis is to obtain the measurement time-series $z(t)$ and then reconstruct the Hankel matrix $\mathcal{H}$ under the assumption that the conditions of the Taken's embedding theorem are satisfied. In discrete case, the measurement time-series can be denoted as $\{z_n\}_{n=1}^N$. Next, we compute the eigen-time-delay coordinates from $\{z_n\}_{n=1}^N$ by taking the SVD of the following Hankel matrix $\mathcal{H}$ or trajectory matrix $\mathcal{D}$ in singular spectrum analysis (SSA) (Wei, 2006) as:

$$\mathcal{H} = \begin{bmatrix} z(t_1) & z(t_2) & \cdots & z(t_p) \\ z(t_2) & z(t_3) & \cdots & z(t_{p+1}) \\ \vdots & \vdots & \ddots & \vdots \\ z(t_q) & z(t_{q+1}) & \cdots & z(t_T) \end{bmatrix} = U\Sigma V^* \tag{9}$$

Here, the time points $t_{i+1} = t_i + \tau$ ($i = 1, \ldots T-1$), $q$ is the number of points in the trajectory and $p$ is the window length. Using the terms in the embedding method with $\tau$ as the time lag and $p$ the embedding dimension, the window length $p$ determines the extent to which we obtain a refined decomposition into basic components and therefore improve the separability of dynamics. In other words, the window length $p$ defines the longest periodicity captured by the Hankel matrix in the HAVOK model. The columns of $U$ and $V$ are arranged hierarchically corresponding to the descending order of singular values in $\Sigma$. In addition, $\mathcal{H}$ often admits a low-rank approximation by the truncation of the first $r$ columns of $U$ and $V$. The low-rank approximation to $\mathcal{H}$ in Eq. (9) gives rise to a measurement subspace that is approximately invariant to $\mathcal{K}$ for the states. Consequently, we can rewrite Eq. (9) with the Koopman operator $\mathcal{K}$:

$$\mathcal{H} = \begin{bmatrix} z(t_1) & \mathcal{K}z(t_2) & \cdots & \mathcal{K}^{p-1}z(t_p) \\ \mathcal{K}z(t_2) & \mathcal{K}^2z(t_3) & \cdots & \mathcal{K}^p z(t_{p+1}) \\ \vdots & \vdots & \ddots & \vdots \\ \mathcal{K}^{q-1}z(t_q) & \mathcal{K}^q z(t_{q+1}) & \cdots & \mathcal{K}^{T-1}z(t_T) \end{bmatrix} \tag{10}$$

The columns and rows of $\mathcal{H}$ in both Eq. (9) and Eq. (10) are well-approximated by the first $r$ truncated columns and rows of $U$ and $V$ respectively, which are called eigen time series providing a Koopman-invariant measurement system.

*2.3.2. Forced linear dynamical system representation*

We can consider the rows of $V$ as a set of coordinates to construct a linear dynamical system. However, a linear model cannot fully capture multiple fixed points, periodic orbits, and the unpredictable chaos with a positive Lyapunov exponent (Rozenbaum, Ganeshan, & Galitski, 2017). To overcome it, the forced linear system has been proposed after applying dynamic mode decomposition (DMD) (Brunton, Proctor, et al., 2016) algorithm to the delay coordinates and obtaining an excellent linear fit for the first $r-1$ variables but bad fit for $v_r$. Particularly, the connection between the eigen-time-delay coordinates and the Koopman operator $\mathcal{K}$ as shown in Eq. (10) induce a linear regression model, in which we build a linear model on the first $r-1$ variables in $V$ and consider $v_r$ as an intermittent forcing term as follows:

$$\frac{d}{dt}\boldsymbol{v}(t) = \boldsymbol{A}\boldsymbol{v}(t) + \boldsymbol{B}v_r(t) = \frac{d}{dt}\begin{bmatrix} v_1 \\ v_2 \\ \vdots \\ v_{r-1} \\ v_r \end{bmatrix} = \begin{bmatrix} \boldsymbol{A} & \boldsymbol{B} \\ \boldsymbol{0} & 0 \end{bmatrix} \begin{bmatrix} v_1 \\ v_2 \\ \vdots \\ v_{r-1} \\ v_r \end{bmatrix} \quad (11)$$

where $\boldsymbol{v} = [v_1 \; v_2 \; \cdots \; v_{r-1}]^T$ is a vector of the first $r-1$ eigen time-delay coordinates. In principle, we can separate the eigen time-delay variables into $r-s$ high-energy modes for the linear model and $s$ low-energy intermittent forcing modes if $v_r(t)$ is not sufficient to model the intermittent forcing. The partition of nonlinear dynamics into deterministic linear dynamics and chaotic dynamics was proposed in the paper of Mezić (Mezić, 2005). The truncated rank $r$ can be estimated by the optimal hard threshold for singular values (Donoho & Gavish, 2013). The HAVOK model extends the dynamics splitting concept to fully chaotic systems, in which the Koopman operators have continuous spectra. We can estimate the matrices $\boldsymbol{A}$ and $\boldsymbol{B}$ in Eq. (11) as follows:

$$\boldsymbol{V}_r = \begin{bmatrix} \boldsymbol{v}^{(1)} & \boldsymbol{v}^{(2)} & \cdots & \boldsymbol{v}^{(q-1)} \\ v_r^{(1)} & v_r^{(2)} & \cdots & v_r^{(q-1)} \end{bmatrix}, \boldsymbol{V}_r' = \begin{bmatrix} \boldsymbol{v}^{(2)} & \boldsymbol{v}^{(3)} & \cdots & \boldsymbol{v}^{(q)} \\ v_r^{(2)} & v_r^{(3)} & \cdots & v_r^{(q)} \end{bmatrix}, \mathcal{A} \approx \boldsymbol{V}_r' \boldsymbol{V}_r^\dagger$$

$$= \begin{bmatrix} \boldsymbol{A} & \boldsymbol{B} \\ \boldsymbol{a} & b \end{bmatrix} \cong \begin{bmatrix} \boldsymbol{A} & \boldsymbol{B} \\ \boldsymbol{0} & 0 \end{bmatrix} \quad (12)$$

In Eq. (12), $\boldsymbol{V}_r'$ is the 1-step time advanced eigen time-delay coordinates of $\boldsymbol{V}_r$. These matrices $\boldsymbol{V}_r$ and $\boldsymbol{V}_r'$ are be related by a best-fit linear operator $\mathcal{A}$ that minimizes the Frobenius norm error $\|\boldsymbol{V}_r' - \mathcal{A}\boldsymbol{V}_r\|_F$, and $\boldsymbol{V}_r^\dagger$ is the pseudo-inverse computed via the

SVD of $V_r$. The estimates $a$ and $b$ are considered bad fit for $v_r$ and approximate to zero. For complex systems of relatively large dimension, the operator $\mathcal{A}$ is also large; therefore, the DMD method or more advanced methods (e.g., sparse identification of nonlinear dynamical systems (SINDy) method (Brunton, Proctor, et al., 2016)) can be applied to consider only the leading eigen-decomposition of $\mathcal{A}$. The HAVOK model's detailed steps are illustrated in Fig. 3.

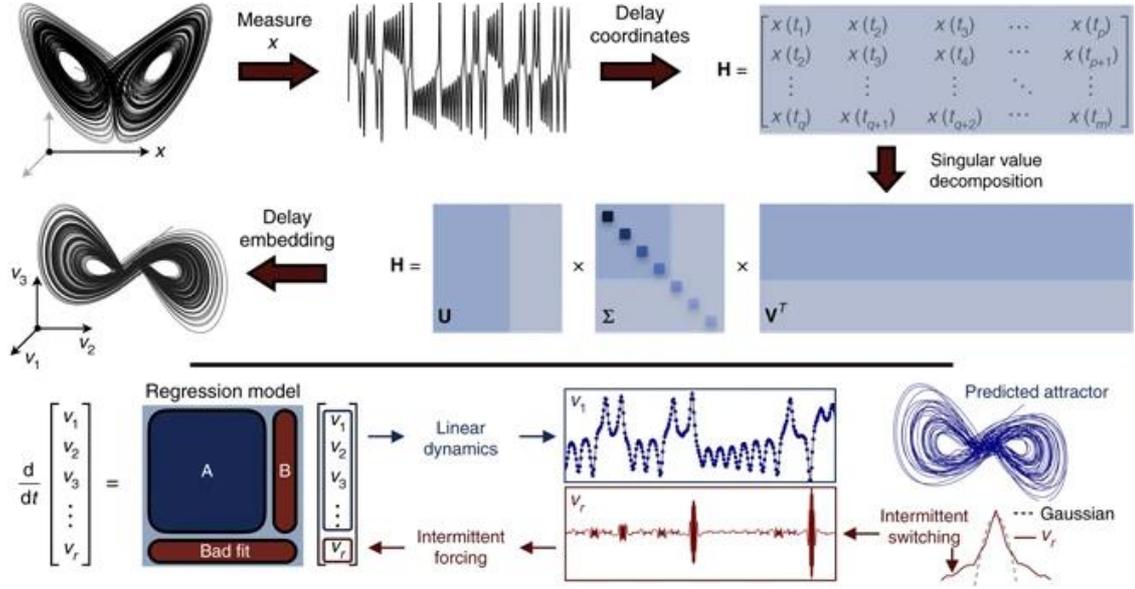

**Figure 2.** Decomposition of chaotic dynamics into linear dynamics with intermittent forcing term. A time series $z(t)$ is used to reconstruct the Hankel matrix $\mathcal{H}$. The singular value decomposition of $\mathcal{H}$ can obtain time-delay coordinates that form a delay-embedded attractor. A best-fit linear dynamical model is built on the first $r-1$ delay coordinates, and the last coordinate $v_r(t)$ is considered a stochastic input that intermittently forces the first $r-1$ variables. The figure is adopted from the HAVOK paper (Brunton et al., 2017).

## 3. Materials and methods

### 3.1. Overview of the methods

The block diagram of the proposed methods is illustrated in Fig. 2. The first step is to perform the HAVOK analysis consisting of two sub-steps: (1) state space reconstruction (see Sub-section 2.3.1) and (2) forced linear dynamical system representation (see Sub-section 2.3.2). Particularly, a state space is reconstructed from the Hankel matrix to obtain eigen time-delay coordinates using the physiological measurement time series $z(t)$. Consequently, a forced linear system using the eigen coordinates was built to model the linear and chaotic intermittent dynamics. The second step is to perform the intermittency analysis proposed by our paper, which comprises two steps: (1) intermittent phases and chaotic bursts analysis (see Sub-section 3.3) and (2) spectral analysis and wavelet analysis of the intermittent forcing component.

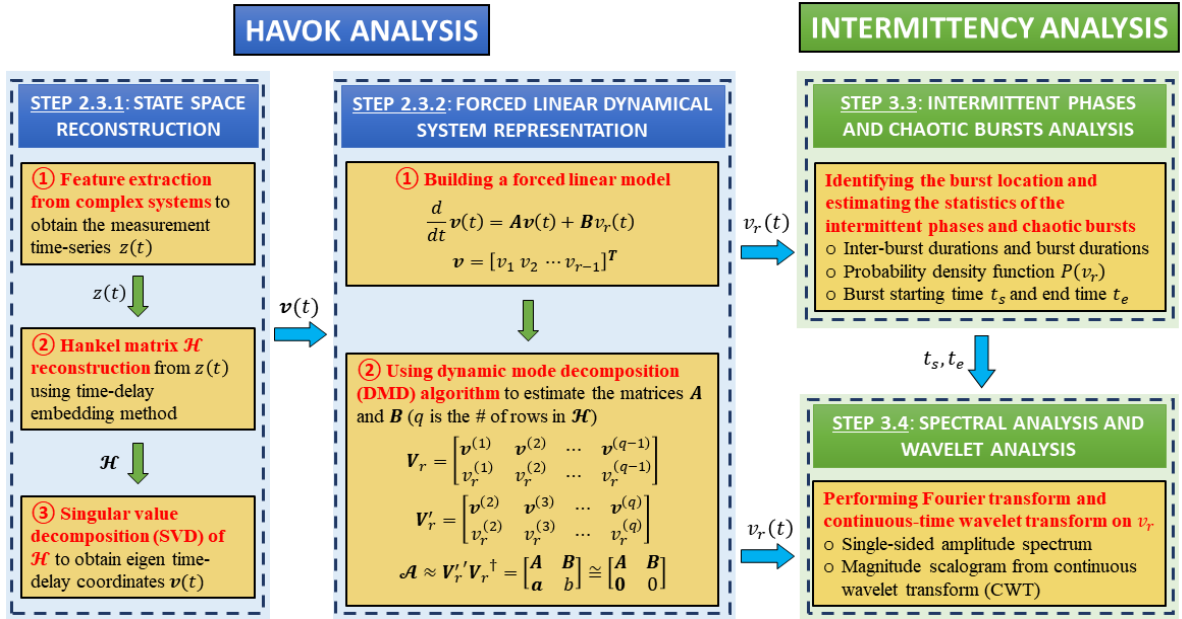

**Figure 3.** Block diagram of the proposed methods which include the performance of the HAVOK analysis and intermittency analysis.

### *3.2. Estimating the statistics of intermittent phases and chaotic bursts:*

In the original work of HAVOK analysis, the method has been applied to many nonlinear dynamical systems including analytical systems, stochastic magnetic field reversal, and real-world systems. The distribution of the intermittent forcing $v_r(t)$ in those examples was shown to be nearly symmetric with fat tails, i.e., the distributions are non-Gaussian. We considered the distribution of the burst durations (denoted as $T_b(k) \in (0, \infty)$), where $k = 1, \ldots, N_b$ is the burst index. First, a hard threshold for detecting active forcing (i.e., burst) is selected such that if $v_r^2(t) \geq \psi \max\{v_r^2(t)|t \geq 0\}$ then the forcing is active, where $\psi$ is a parameter. The $\psi$ parameter can be fine-tuned to achieve the best agreement with the intermittent behaviors of the system in the training data, e.g., lobe switching in Lorenz system or the disease onsets in pathophysiological processes. Second, we define the starting and ending time of the bursts as $t_s(k)$ and $t_e(k)$ and burst starting-ending time as $T_b(k) := t_e(k) - t_s(k)$. From the notations, we can also define the inter-burst duration as $T_{ib} = t_s(k) - t_e(k-1)$ for $k = 2, \ldots N_b$. Those statistics of intermittent phases and chaotic bursts are illustrated in Fig. 4. The distributions of $v_r, T_b,$ and $T_{ib}$ can be estimated using non-parametric histogram-based pdf estimation method (Fahmy, 2010).

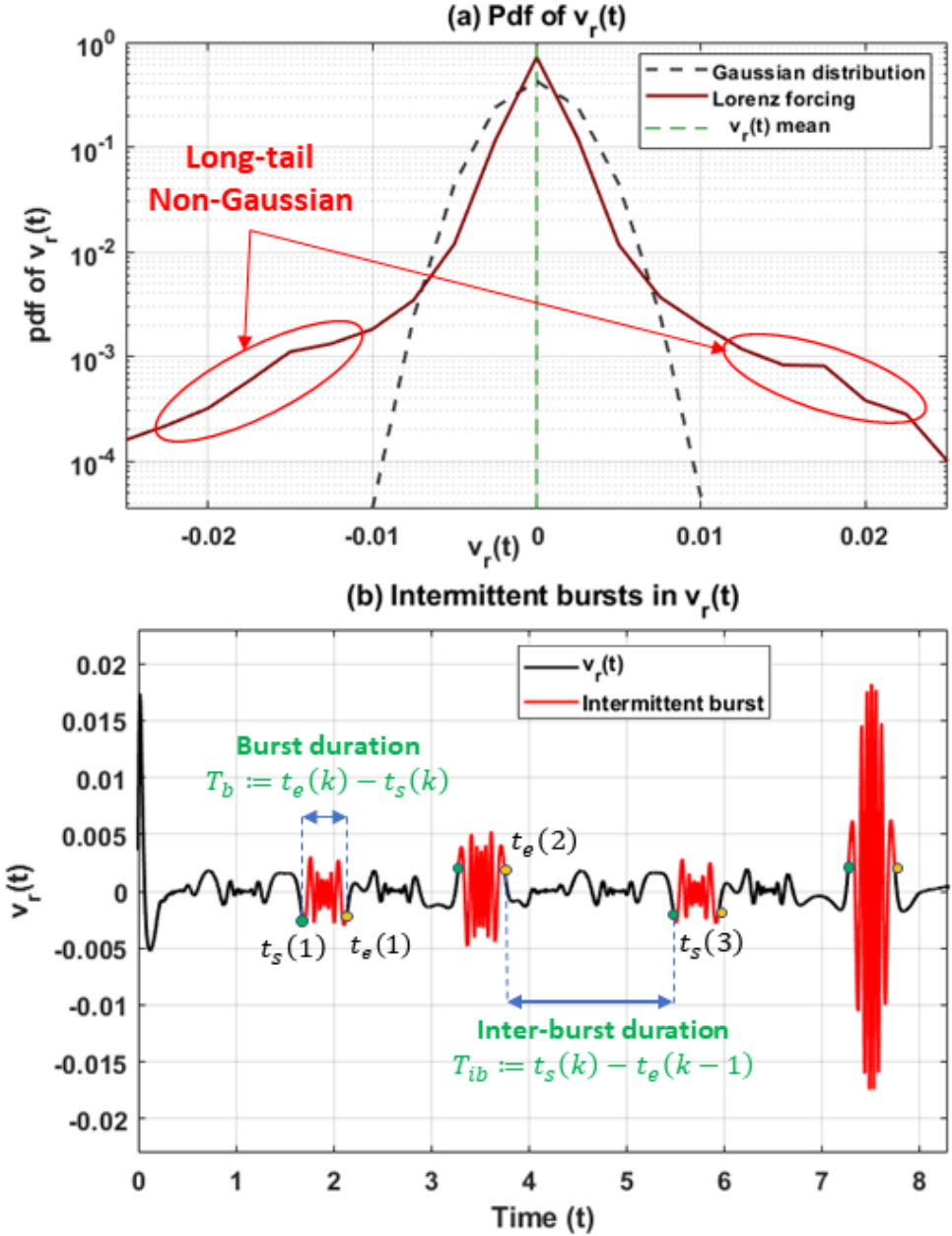

**Figure 4.** Illustration of the statistics of intermittent phases and chaotic bursts. Panel (a) shows the long-tail non-gaussian distribution of the forcing term $v_r(t)$ for the Lorenz system. Panel (b) illustrates the estimation of burst duration $T_b$ and inter-burst duration $T_{ib}$ in $v_r(t)$, where $t_s(k)$ and $t_e(k)$ are the starting time and ending time of the bursts.

### 3.3. Spectral analysis and wavelet analysis of the intermittent forcing $v_r$

The statistics of the intermittent forcing are not sufficient to characterize the mode switching of nonlinear dynamics because the burst timing is also essential information. Furthermore, the forcing corresponding to the long tails of the distribution comes in the form high-frequency bursts, which cannot be captured using the statistics of intermittent phases and chaotic bursts alone. To better understand the characteristics of $v_r$, a fast

Fourier transform (FFT) (Oppenheim et al., 1992) can be applied to $v_r$ to convert the forcing signal from the time domain to a representation in the frequency domain. This is performed by decomposing $v_r$ into components of different frequencies as:

$$\boldsymbol{V} = \mathcal{F}(v_r), \qquad V_k = \sum_{t=0}^{q-1} v_r^{(t)} \cdot e^{-\frac{i2\pi}{q}kt} \tag{13}$$

where the Fourier transform operator is denoted by $\mathcal{F}(\cdot)$, $q$ is the number of rows in the Hankel matrix $\boldsymbol{\mathcal{H}}$ and $k$ is the frequency. In our spectral analysis on $v_r$, we utilize a plot tool called single-sided (positive frequencies) amplitude spectrum, where it computes the amplitude at each positive frequency, i.e., the amplitude of each sinusoidal component in the FFT analysis, and the maximum frequency is equal to half of the sampling frequency $f_s$. We also performed Continuous Wavelet Transform (CWT) (Lilly, 2017) on $v_r$ to obtain an overcomplete representation of $v_r$ in time-frequency domain. The CWT of the forcing signal $v_r(t)$ at a scale ($a > 0$) and translational value $b \in \mathbb{R}$:

$$X_w(a,b) = \frac{1}{|a|^{1/2}} \int_{-\infty}^{\infty} v_r(t) \bar{\psi}\left(\frac{t-b}{a}\right) dt \tag{14}$$

where $\psi(t)$ is a continuous function called the mother wavelet (e.g., Morse wavelet) and $\bar{\psi}$ is its complex conjugate. The main objective of the mother wavelet is to provide a function to generate the translated and scaled versions (called "daughter" wavelets) of the mother wavelet. To display the results of CWT, the scalogram is a useful tool to represent the absolute value of the CWT of a signal, which is plotted as a function of frequency and time. For the intermittency analysis, the scalogram is a well-suited method for analyzing the intermittent forcing $v_r(t)$ that occurs at different frequency scales. The scalogram computes the modulus of CWT coefficients (i.e., $|X_w|$). Therefore, we can obtain time-localization for short-duration, high-frequency bursting events, and better separate them with low-frequency components and longer-duration events.

### 3.4. Apnea-ECG Database for OSA

We performed proposed intermittent forcing analysis on the pathophysiological processes of Obstructive Sleep Apnea (OSA) – a common sleep breathing disorder with long-term effects on the cardiorespiratory system. We have selected the Apnea-ECG Database (Penzel, Moody, Mark, Goldberger, & Peter, 2000) from Physionet.org to investigate the nonlinear intermittent behaviors. The data set has 70 records of OSA patients, and the recordings vary from 7 hours to nearly 10 hours each. Each recording comprises of continuous digitized ECG signal and apnea-hypopnea event expert-labeled annotations.

The annotations consist of the binary-coded events: "Normal breathing" (N) or "Disordered breathing" (A) for each minute of the recording. The disordered breathing event may correspond to a single apnea episode or a hypopnea episode or may include a longer sequence of apneas and hypopneas. Consequently, 26 records were shortlisted to be included in our HAVOK and intermittency analysis for OSA since we do not consider healthy (normal) subjects and patients with very mild (AHI < 5) or severe sleep apnea (AHI > 30) with long-duration apnea-hypopnea events, where AHI is apnea-hypopnea index that indicates the number of apneas or hypopneas documented during the sleep study per hour. This is because the severe apnea cases with consistent apnea-hypopnea events result in the absence of the intermittent switching between apnea-hypopnea events and normal episodes.

## 4. Results

The case study of the pathophysiological processes in Obstructive Sleep Apnea (OSA) is investigated, in which we performed four main steps: (1) signal processing and feature extraction, (2) state space reconstruction, (3) HAVOK analysis for the OSA cases study, and (4) performing intermittent forcing analysis.

### *4.1. Signal processing and feature extraction*

The signal preprocessing and feature extraction were performed. At first, a $5^{th}$ order Butterworth 0.5–30 Hz bandpass filter has been applied to the signal to eliminate noises and baseline wandering. Subsequently, the Hamilton-Tompkins algorithm (Wu, Zheng, Chu, & He, 2020) was employed to detect R peaks and compute RR intervals. A set of 18 features was extracted from Heart Rate Variability (HRV) Tool (Vollmer, 2019), this is an open-source MATLAB Toolbox that quantifies the spectral energy and nonlinear patterns of the HRV signals. Heart rate variability (HRV) features (Pecchia, Castaldo, Montesinos, & Melillo, 2018) quantify the fluctuation in the time intervals between adjacent heartbeats, which arise from the neurocardiac functions, the heart-brain interactions, and dynamic non-linear autonomic nervous system (ANS) processes. Previous studies have shown that power spectral features extracted from HRV signals are clinically significant for obstructive sleep apnea detection and prediction (Atri & Mohebbi, 2015; Hossen, 2015; Le & Bukkapatnam, 2016; Urtnasan, Park, Joo, & Lee, 2018). Next, the sliding window length is selected to be one minute according to the 1-minute OSA annotations. Detailed steps can be found in our previous works (Le, Cheng, Sangasoongsong, Wongdhamma, & Bukkapatnam, 2013).

*4.2. State space reconstruction and eigen time-delay coordinates*

First, we constructed the Hankel matrix $\mathcal{H}$ of the size $q \times p$ from a measurement (i.e., extracted HRV feature) time series, where $q = N - p + 1$, $N$ is the length of the feature time series, and $p$ is the window length as introduced in Sub-Section 3.2. The window size $p$ was chosen to be 15 according to our previous study (Le et al., 2013) such that the maximum periodicity of the underlying process can be captured. We selected the most appropriate HRV feature, namely the TRI feature, to construct the Hankel matrix and obtain eigen-time-delay coordinates $v(t)$. The TRI feature computes the triangular index from the RR-interval histogram, which is an essential metric to quantify the heart rate variability (Byun et al., 2019). The feature was selected so that the abnormal (apneic) and normal states are linearly or potentially nonlinear separable (not mixing) in the reconstructed state space with a strong correlation between the active forcing signature and apneic events. The reconstructed state spaces of 4 patients were illustrated in Fig. 5.

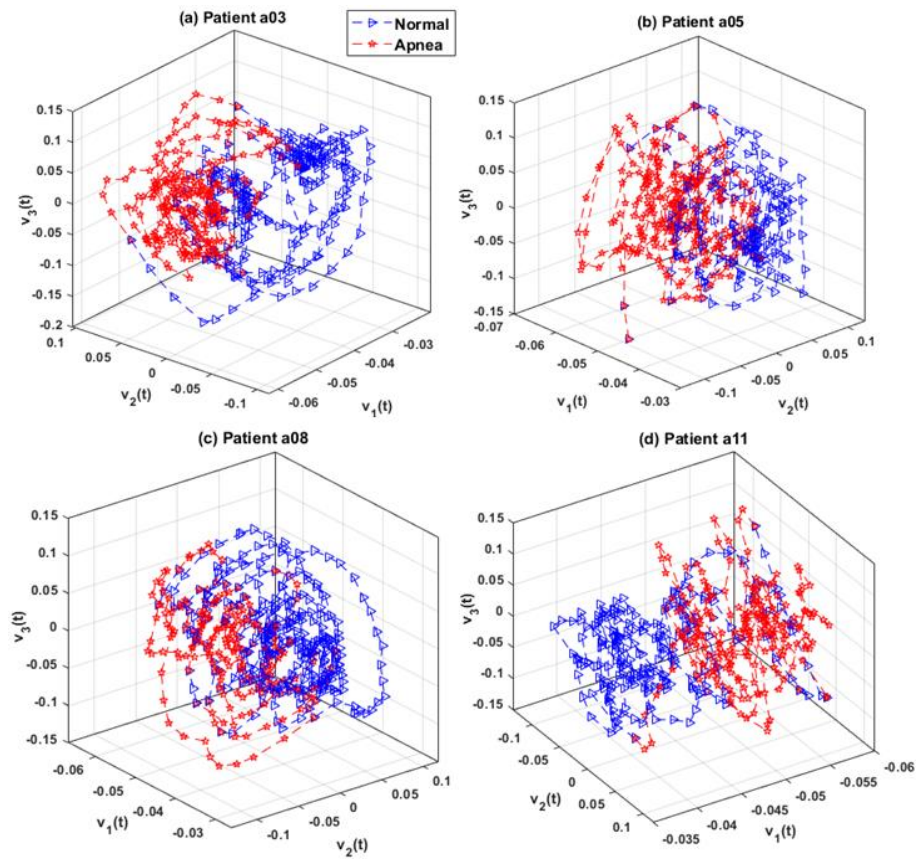

**Figure 5.** The embedded state spaces of 4 representative patients, in which $v_1$, $v_2$, and $v_3$ are the first three eigen time-delay coordinates obtained from Hankel matrix using the TRI feature. The reconstructed state spaces of the representative patients a03, a05, a08, and a11 are illustrated in panel (a), (b), (c), and (d), respectively.

## 4.3. HAVOK analysis for OSA

In continue, a Koopman linear system on the eigen time-delay coordinates with forcing term was reconstructed to model both linear and chaotic intermittent dynamics of OSA. Here, the truncated rank $r$ was selected using optimal hard threshold for singular values (Donoho & Gavish, 2013). After performing the HAVOK analysis, we obtained the intermittent forcing $v_r$ with a hard threshold determined by $\max v_r^2$ for active input forcing and illustrated the results for one representative patient a03 in Fig. 6.

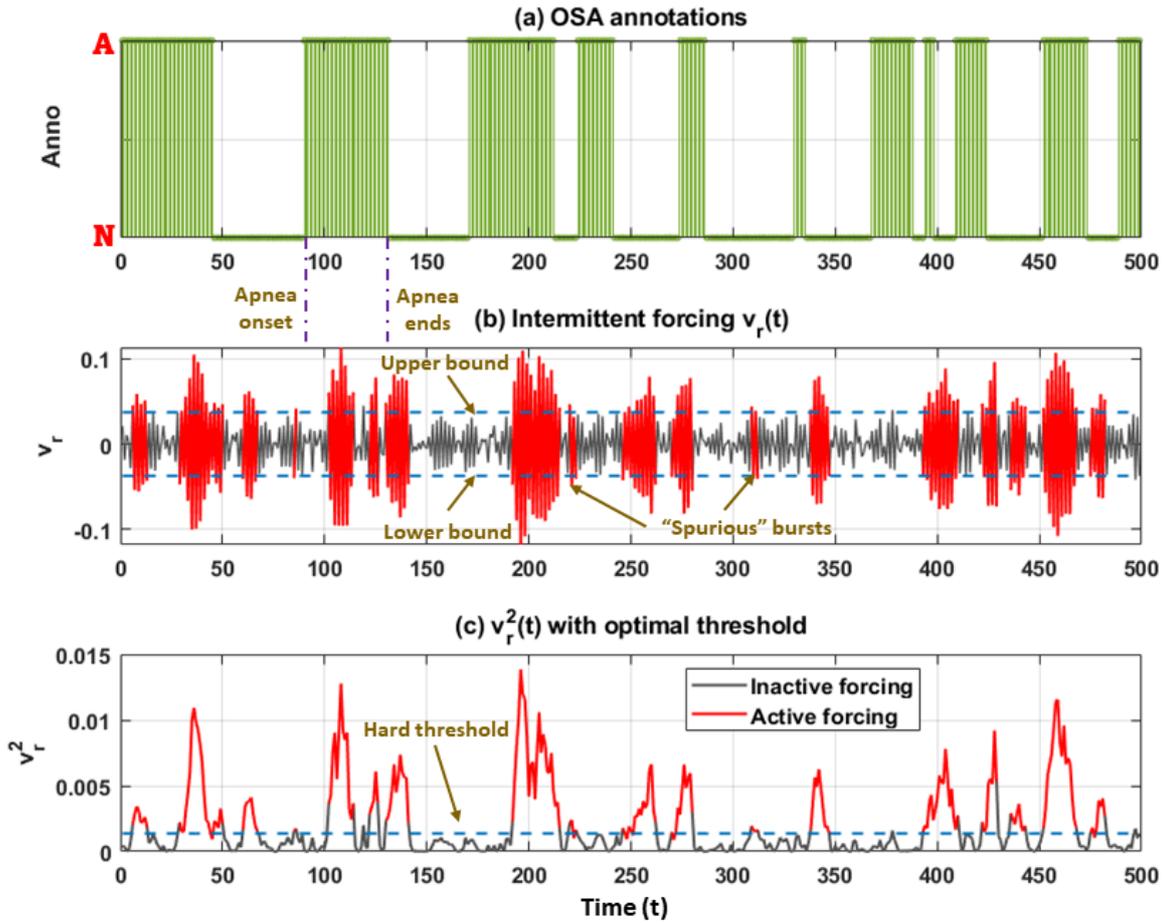

**Figure 6.** Forcing term $v_r$ with hard threshold determined by $\max v_r^2$ of patient a03. When the forcing is active (indicated by red line), it corresponds to the period over which $v_r^2(t) \geq \psi \max\{v_r^2(t)|t \geq 0\}$ ($\psi = 0.12$). The active forcing represents the intermittent behaviors (possibly chaotic) before, during, and after an apneic episode, which can be shown by a strong association between the OSA annotations ("N" is "normal breathing" and "A" is "apnea-hypopnea event") and the timing of the bursts. The characteristics of the intermittent bursts are high-amplitude, high-frequency, and usually associated with apnea-hypopnea events. However, due to the limitation of the hard threshold, some

"spurious" bursts were observed, which might not be the actual intermittency from the OSA pathophysiological processes.

### 4.4. Intermittent forcing $v_r$ analysis

To analyze and understand the intermittent component $v_r$, we estimated the statistics of intermittent phases and chaotic bursts and performed spectral and wavelet analysis on $v_r$. Firstly, we approximated the distribution of $v_r$ using the histogram-based probability density function estimation method. The distributions of $v_r(t)$ for 4 patients a03, a05, a08, and a11 were represented in Fig. 7.

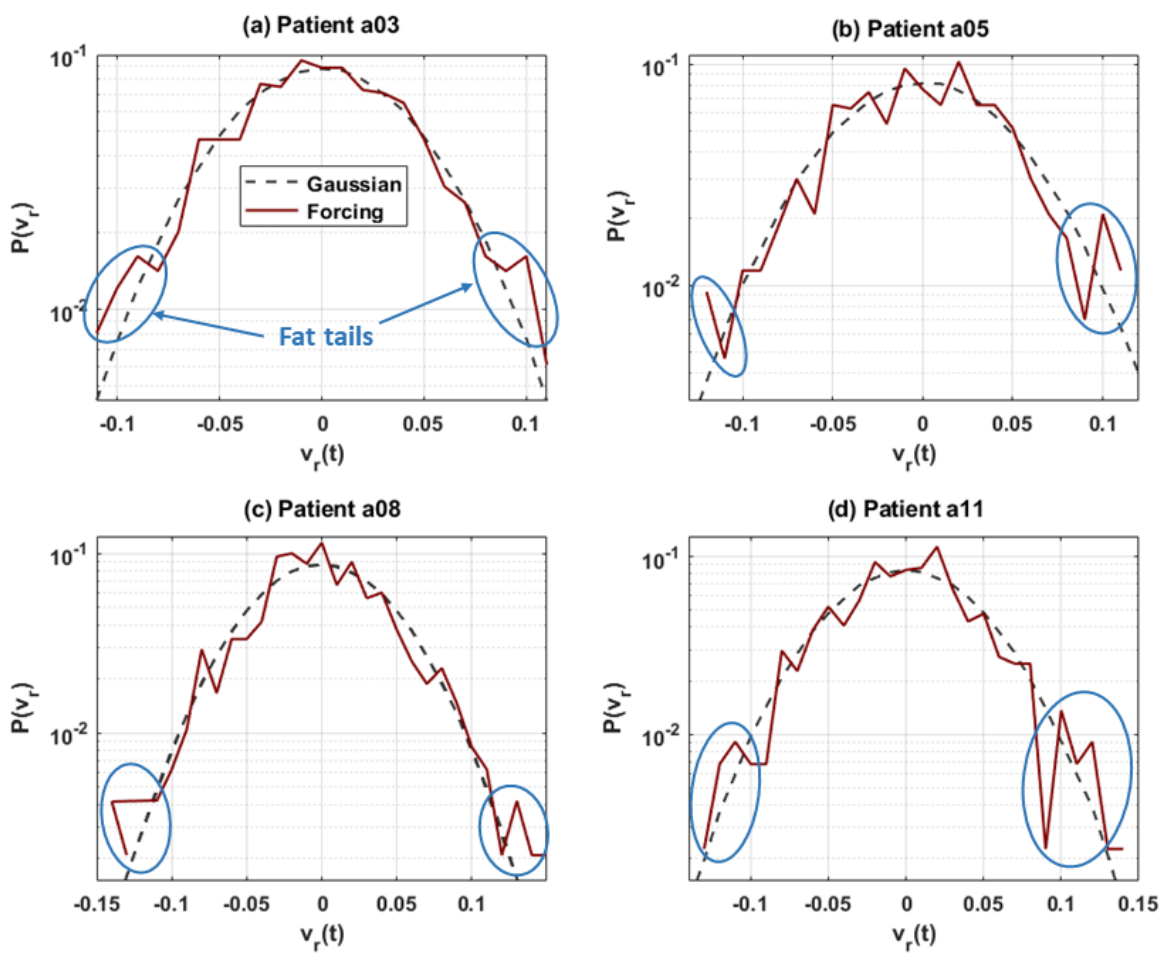

**Figure 7.** Fat-tailed distribution of the forcing term $v_r(t)$ of patient a03, a05, a08, and a11. The distributions of those 4 patients were nearly symmetrical and aligned with the Gaussian distribution curve well except for the tails of the distribution. The "fat tails" appeared to arise from the high-amplitude and high-frequency bursting in $v_r(t)$.

Because the statistics of the intermittent forcing are not sufficient to characterize the intermittency of OSA nonlinear dynamics and the forcing corresponding to the long tails of the distribution comes in the form of high-frequency bursts, we performed fast Fourier transform (FFT) and Continuous Wavelet Transform (CWT) on $v_r(t)$ on 1 representative patient a03, which was shown in Fig. 8. For the CWT analysis, we selected to mother wavelet to be the analytic Morse wavelet, whose shape is similar to the bursting pattern. A scalogram of the intermittent forcing $v_r$ was generated to represent the change in the modulus of CWT amplitude $|X_w|$ over time and across different frequencies. Based on the results of the scalogram, we investigated the intermittent forcing energy distribution over time using the FFT analysis. Specifically, we estimated the single-sided amplitude spectrum of the complete time series $v_r$, the dominant bandwidth $B \coloneqq f_L - f_H$ ($f_L$ and $f_H$ are the low and high-frequency limits of the dominant bandwidth) was also calculated, which accounts for 95% energy of the signal. Next, we selected three windows of interest, namely $W_1$, $W_2$, and $W_3$, in which the magnitude of the scalogram was high and they correspond to a sequence of apneic events. The amplitude spectra for those 3 windows were consequently estimated to characterize the intermittency phenomenon in those windows as shown in Fig. 9.

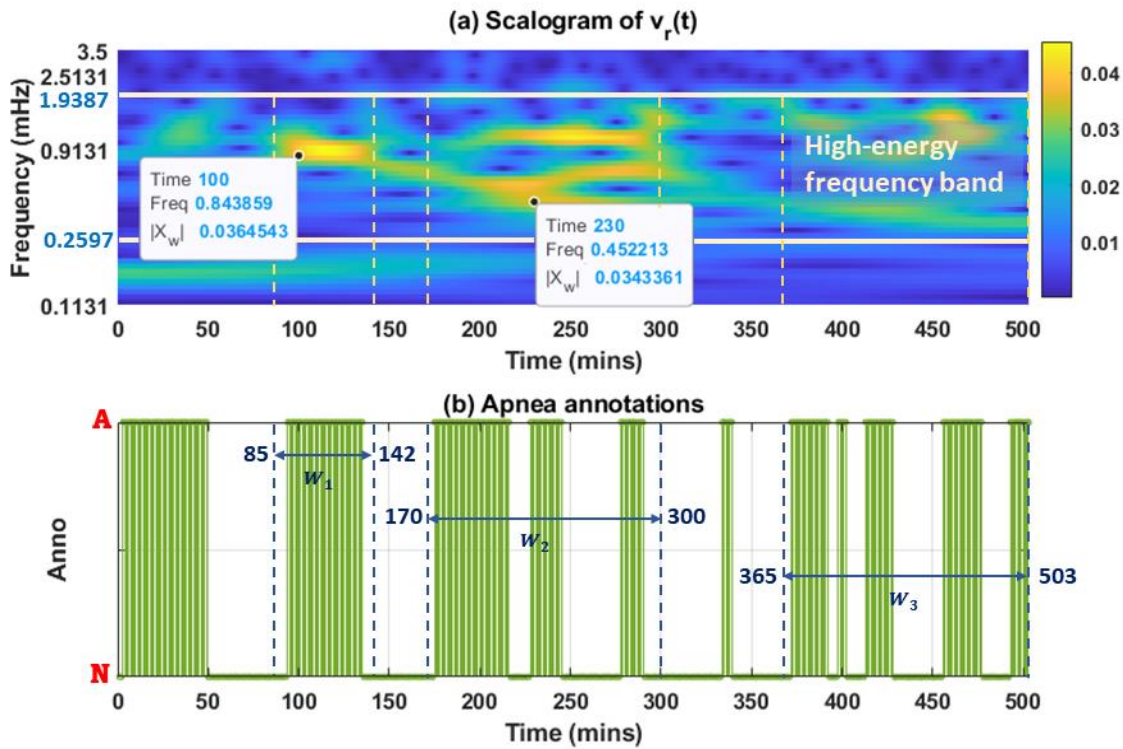

**Figure 8.** Scalogram of the intermittent forcing $v_r$ with magnitude in frequency-time domain after CWT (a) and the OSA annotations of patient a03 scored by the experts (b). As shown in the scalogram, the most representative scales or frequencies of the forcing are from 0.26 mHz to 1.94 mHz, the energy captured by the frequencies belonging to this bandwidth is 95%. The time points at which the power $|X_w|$ had high values were highly associated with the apnea-hypopnea onset timing ("N" is "normal breathing" and "A" is "apnea-hypopnea event") as shown in panel (b), e.g., the modulus $|X_w| \approx 0.0365$ at minute 100 or $|X_w| \approx 0.0343$ at minute 230. In panel (b), three windows $W_1$, $W_2$, and $W_3$ corresponding to the apneic events were selected to investigate the intermittency of OSA using spectral analysis FFT during the high-magnitude periods in the scalogram.

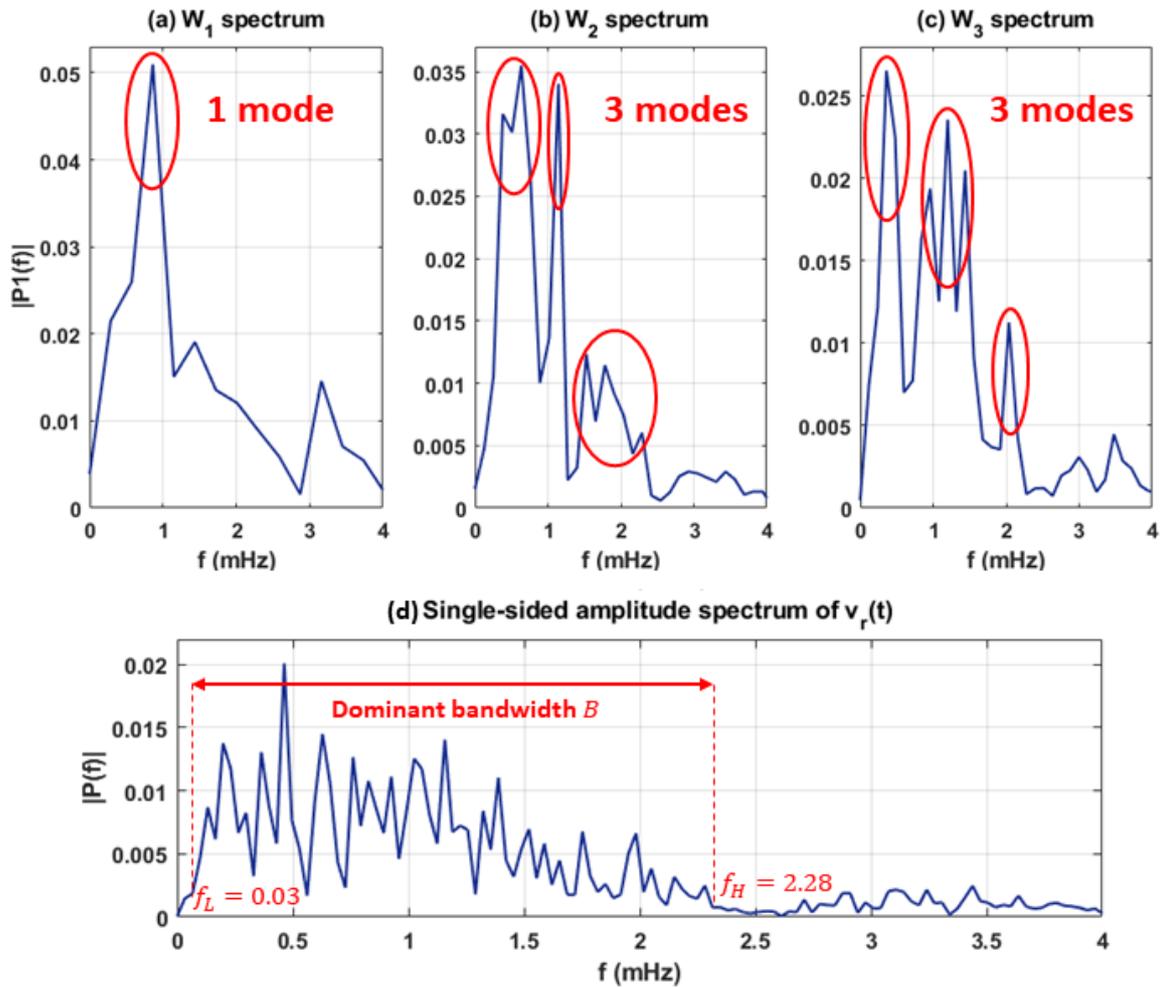

**Figure 9.** Single-sided amplitude spectrum of the windows $W_1$, $W_2$, $W_3$, and the forcing $v_r$ of patient a03. Panel (a), (b), (c): the amplitude spectra of the signals in the windows $W_1$, $W_2$, and $W_3$ after performing FFT. In the spectrum of $W_1$, there was 1 predominant frequency peaked at 0.86 mHz, which can be interpreted as 1 mode of intermittency. In contrast, there were 3 modes in the spectra of $W_2$ and $W_3$. For $W_2$, the 1st mode peaked at two frequencies 0.38 mHz and 0.64 mHz, the 2nd mode peaked at 1.15 mHz, and the last mode also peaked at two frequencies 1.53 mHz and 1.78 mHz. Regarding $W_3$ spectral analysis, the peak of the 1st mode was at 0.36 mHz, the 2nd mode had 3 predominant frequencies peaked at 0.96 mHz, 1.20 mHz, and 1.44 mHz, and the last mode peaked at

2.04 mHz. Panel (d): the single-sided amplitude spectrum of $v_r$ with large amplitude in the frequency bandwidth $B$, which ranges from 0.03 to 2.28 mHz.

We performed the HAVOK analysis and the intermittent forcing analysis on 26 OSA patients. The results were summarized in Table 1.

Table 1. Summary of the HAVOK analysis and the intermittent forcing analysis for 26 OSA patients. Here, we reported the patient ID, their corresponding AHI, the truncated rank $r$, the energy captured by intermittent forcing $v_r$, the burst duration $T_b$, the inter-burst duration $T_{ib}$, and the dominant bandwidth $B := f_L - f_H$. The overall statistics were also estimated for each column of the table.

| Patient ID | AHI | Rank $r$ | $v_r$ energy (%) | $T_b$ in mins (mean $\pm$ SD) | $T_{ib}$ in mins (mean $\pm$ SD) | Dominant $B$ (mHz) |
|---|---|---|---|---|---|---|
| a03 | 39.1 | 10 | 3.69 | 2.09 $\pm$ 1.90 | 12.30 $\pm$ 12.86 | .03 - 2.28 |
| a05 | 41 | 8 | 3.96 | 2.29 $\pm$ 3.39 | 10.16 $\pm$ 7.57 | .08 - 2.87 |
| a06 | 24.7 | 10 | 3.32 | 1.52 $\pm$ 1.31 | 15.92 $\pm$ 17.15 | .06 - 4.32 |
| a07 | 63 | 9 | 4.08 | 1.69 $\pm$ 1.58 | 10.66 $\pm$ 8.58 | .07 - 2.51 |
| a08 | 42 | 9 | 4.40 | 1.96 $\pm$ 1.99 | 10.74 $\pm$ 10.65 | .07 - 2.48 |
| a11 | 14 | 8 | 4.41 | 1.79 $\pm$ 1.88 | 9.62 $\pm$ 7.06 | .04 - 2.87 |
| a13 | 42 | 9 | 4.22 | 1.69 $\pm$ 1.44 | 9.37 $\pm$ 9.00 | .03 - 2.63 |
| a15 | 52 | 9 | 4.25 | 1.82 $\pm$ 1.39 | 8.47 $\pm$ 7.04 | .07 - 2.65 |
| a19 | 34 | 9 | 3.91 | 1.38 $\pm$ 0.90 | 12.49 $\pm$ 11.71 | .07 - 2.45 |
| a20 | 41 | 9 | 4.20 | 2.10 $\pm$ 1.84 | 16.88 $\pm$ 14.43 | .03 - 2.78 |
| b03 | 24 | 9 | 4.53 | 1.63 $\pm$ 1.40 | 16.23 $\pm$ 12.49 | .04 - 2.76 |
| x02 | 37.7 | 8 | 4.64 | 1.65 $\pm$ 0.74 | 8.55 $\pm$ 8.61 | .04 - 2.84 |
| x05 | 34 | 8 | 4.38 | 1.68 $\pm$ 1.34 | 15.62 $\pm$ 12.78 | .07 - 2.76 .15 - 7.11 |
| x07 | 21 | 10 | 4.10 | 1.84 $\pm$ 1.53 | 13.80 $\pm$ 13.24 | .10 - 2.21 .06 - 7.31 |
| x08 | 48 | 10 | 4.22 | 1.44 $\pm$ 1.10 | 7.46 $\pm$ 8.39 | .03 - 2.05 |
| x09 | 18.5 | 11 | 4.22 | 1.83 $\pm$ 2.99 | 10.62 $\pm$ 12.78 | .06 - 2.07 |
| x10 | 10 | 10 | 3.88 | 1.46 $\pm$ 1.17 | 12.77 $\pm$ 19.49 | .07 - 2.61 |
| x12 | 33 | 10 | 4.00 | 1.37 $\pm$ 0.93 | 8.31 $\pm$ 9.12 | .06 - 2.09 |
| x13 | 18.7 | 12 | 3.88 | 1.60 ± 1.84 | 14.79 ± 18.22 | .03 - 2.00 |
| x15 | 15.9 | 9 | 4.52 | 1.68 ± 1.72 | 14.06 ± 14.91 | .10 - 2.55 |
| x16 | 24 | 9 | 4.30 | 1.45 ± 1.15 | 10.44 ± 8.88 | .06 - 2.57 |
| x19 | 56.2 | 10 | 4.20 | 2.31 ± 2.69 | 10.42 ± 8.26 | .03 - 2.12 |
| x21 | 19 | 9 | 4.14 | 1.47 ± 1.21 | 13.00 ± 13.54 | .03 - 2.64 |
| x23 | 14.3 | 10 | 4.36 | 1.53 ± 1.42 | 10.57 ± 9.70 | .10 - 3.74 |
| x25 | 48 | 9 | 4.26 | 1.96 ± 2.18 | 9.00 ± 7.86 | .06 - 2.63 |
| x26 | 15.1 | 9 | 4.86 | 1.66 ± 1.66 | 12.23 ± 16.80 | .10-2.81 |
| Overall statistics | 31.93 ± 14.71 | 9.35 ± 0.94 | 4.19 ± 0.31 | 1.73 ± 0.26 | 11.71 ± 2.69 | .09 ± .02 – 2.63 ± 0.50 |

*Overall statistics estimate the mean and standard deviation for each column*

According to Table 1, the estimated truncated rank $r$ from the SVD of the Hankel matrix shown in Eq. (11) varied from $r = 8$ to $r = 12$, and its pooled (inter-patient) mean and standard deviation were $9.35 \pm 0.94$, which showed a consistent rank to reconstruct the dynamics. Here, the model rank was determined mainly from the model accuracy on validation data, i.e., the holdout data set for hyper-parameter tuning, the accuracy of OSA attractor dynamics reconstruction, and the clear association between forcing signature and OSA-related events. In addition, the percentage of energy accounted for by the forcing $v_r(t)$, which corresponds to the low-energy modes of the dynamics, was $4.19\% \pm 0.31$ across patients. The pooled means and standard deviations of the burst duration $T_b$ and the inter-burst duration $T_{ib}$ were also calculated to be $1.73 \pm 0.26$ minutes and $11.71 \pm 2.69$ minutes, respectively. Those estimated pooled statistics had small inter-patient variances; however, the intra-patient variances were significantly high which showed the large variation in the burst and inter-burst durations during the apnea-hypopnea events. To examine the strength and direction of the linear relationship among a group of 3 variables, namely patients' AHI, $T_b$, and $T_{ib}$, we performed the bivariate Pearson Correlation test (Benesty, Chen, Huang, & Cohen, 2009) for two pairs in the group: (1) AHI and $T_b$ and (2) AHI and $T_{ib}$. The sample correlation coefficients were reported to be $r_{AHI,T_b} = 0.4364$ (p-value = 0.0258) for the first pair and $r_{AHI,T_{ib}} = -0.3872$ (p-value = 0.0507) for the latter. The p-value of the Pearson correlation test of the last pair was marginally above the significance level $\alpha = 0.05$ demonstrating that AHI is significantly positively correlated with $T_b$ and negatively correlated with $T_{ib}$, i.e., the higher the value of AHI is, the larger the duration and the occurrence frequency of the bursts are. Lastly, the estimated dominant bandwidths of $v_r$ of all patients in the single-sided amplitude spectrum were quite similar, which can be demonstrated by the small fluctuations in the values of the low and high frequency limits $f_L$ and $f_H$, in which the inter-patient means and standard deviations of $f_L$ and $f_H$ are $0.09 \pm 0.02$ mHz and $2.63 \pm 0.50$ mHz.

## 5. Discussion

We investigated the intermittent nonlinear dynamics of obstructive sleep apnea (OSA) as a case study. In our analysis, the measurement time series obtained from the complex systems poses several problems. Based on the concepts of chaos from nonlinear dynamics theory, the chaotic bursts are considered to belong to an invariant subspace; however, we only have limited information about the characteristics of this invariant subspace because of the rare occurrence of those chaotic bursts. Moreover, over the period of "bursting" events, the time scale of the dynamics is much faster, and the uniform sampling rate of the measurement might be too modest to resolve them for a successful analysis of the intermittent chaotic bursts. Champion et. al. has proposed different sampling strategies that efficiently handle multi-scale systems (Champion, Brunton, & Kutz, 2019), in which the dynamics can be separated into fast and slow dynamics. However, the main goal of our paper is the intermittency dynamics analysis but not all phenomena of the nonlinear dynamics. Hence, we only consider the time scale that is suitable to obtain the intermittent forcing accurately. Furthermore, there are other additional difficulties arising from the inhomogeneity of the time series for the intermittent forcing analysis, e.g., the non-stationary data with intermittent behaviors.

When applying the HAVOK model, the selection of model hyperparameters should be carefully taken into account. First, the number of rows $q$ in the Hankel matrix should be chosen to obtain an appropriate delay embedding basis $U$, and the time delay $\tau$ should be selected based on the embedding dimension $p$ to enforce the independence of the time series. Second, the truncation rank $r$ was selected using mainly the following criteria: (1) the model accuracy on the validation data set, (2) the strong correlation between forcing signature and important observed intermittent events, and (3) the sparse structure of the matrix $A$ in the linear model. For our OSA case study, we selected the number of columns of the Hankel matrix $p = 15$, $\tau = 1$ minutes, and the truncation rank $r$ from $r = 8$ to $r = 12$ using the aforementioned rules. To select the rank $r$ for each OSA patient summarized in Table 1, we increased the rank until we only observe the corrupted statistical noise in the $v_{r+1}(t)$ time series, and we saw a strong correlation between the OSA annotations and the intermittent forcing $v_r(t)$ during the apneic events. The correlation of the forcing signature and the observed chaotic burst shown in Fig. 6 was sensitive to the truncation rank $r$ as we varied this value.

The results from the intermittent forcing analysis have highlighted different interesting features of the intermittency dynamics underlying obstructive sleep apnea. First, the intermittent behaviors exhibit in the form of high-amplitude and high-frequency chaotic bursts. Those are the key characteristics of the intermittent behaviors observed in other nonlinear dynamical systems. Moreover, we have observed a forcing signature associated with a long "disordered breathing" episode (illustrated in Fig. 6), in which was the forcing amplitude of $v_r(t)$ started increasing after the apnea-hypopnea onset and remained high within a short period after its termination. This pattern can even be observed more clearly in the magnitude scalogram of $v_r(t)$ in frequency-time domain after CWT as shown in Fig. 8, where we saw the correlation between the value of the scalogram amplitude $|X_w|$ and the apnea-hypopnea events and the splitting of the intermittent energy into multiple energy bands. About 95% of the energy was captured within the frequency bandwidth from 0.26 mHz to 1.94 mHz, which was consistent with the dominant bandwidth in the spectrum of $v_r$ shown in Fig. 9. To investigate further the distribution of the intermittent energy over time, we selected three windows $W_1$, $W_2$, and $W_3$, and we found there are various predominant frequencies and modes associated with the apneic events. The fluctuations in the high-energy frequencies and the number of modes in each window may correspond to different intermittent behaviors arising from a mixture of hypopneas and different types of apneas: obstructive apnea, central apnea, and complex apnea (Kunisaki et al., 2018). Regarding the statistics of intermittent phases and chaotic bursts. The "fat tails" feature was observed consistently in the shortlisted patients in Fig. 7, which may correspond to the actual intermittency of the underlying process. The estimated pooled statistics of burst and inter-burst duration had small variances, which highlighted the consistency and robustness of the estimates. Furthermore, we observed a significantly strong correlation between the burst and inter-burst durations with the apnea-hypopnea episode durations and apnea-hypopnea inter-arrival time respectively. In addition, after examining the estimated Person correlation coefficients of (1) AHI and $T_b$ and (2) AHI and $T_{ib}$, we concluded the higher the value of AHI is, the larger the duration and the occurrence frequency of the bursts are. This may serve as the evidence to support the hypothesis that the forcing signature arising from the intermittency dynamics of OSA will become more prominent and occur more frequently if the OSA condition exacerbates. Therefore, there is strong evidence that the observed statistics and spectral properties of the intermittent forcing capture well the intermittent dynamics of OSA pathophysiology.

Overall, the analysis and decomposition of the intermittent forcing $v_r$ characterized the spectral properties of $v_r$ and found the most representative bandwidth to represent $v_r$ itself. The selected HRV features reflect the activities and effects of a mixture of both sympathetic and parasympathetic systems, *e.g.,* vagal-mediated modulation of heart rate (Chang et al., 2020). However, HRV features can be insufficient to provide information about respiration-related paroxysmal events such as APN-HYPs, therefore many studies have combined HRV features with ECG-derived respiratory signals (EDR) to improve the richness of features and model accuracy (Atri & Mohebbi, 2015; Mendez, Bianchi, Matteucci, Cerutti, & Penzel, 2009). We hypothesize the representative scales of $v_r$ from the Hankel matrix built from HRV features are associated with the long time-scale dynamics of sympathetic and parasympathetic effects and more investigations on this is necessitated.

## 6. Conclusion

In summary, our paper proposed an intermittency analysis method that systematically decomposes and analyzes the intermittent forcing obtained from the HAVOK model. Particularly, in addition to the distribution of the intermittent forcing, we proposed additional statistics such as burst starting-ending time and burst and inter-burst duration quantifiers to better characterize the intermittent phases and chaotic bursts. Moreover, the temporal spectra of the intermittent forcing signal estimated from the Discrete Fourier transform were used to characterize the spectral properties of the intermittent forcing the adaptive continuous-time wavelet analysis with different types of mother wavelets chosen to match the morphological features of the bursts was performed to extract local spectral and temporal information simultaneously. To validate our proposed methods, a sleep disorder, namely obstructive sleep apnea, was selected as the case study. This is the first attempt to characterizing chaos in pathophysiological processes in a sleep disorder like OSA. In our case study, the forcing signal can predict the intermittent transient events, such as the transition from normal to apneic episodes in the OSA case study. Moreover, the eigen time-delay coordinates were obtained to find the intrinsic states that serve as the simple representation of complex systems thanks to the availability of big data and advanced machine learning algorithms. The intermittent forcing input to the system analysis captures the transient transitions from normal to abnormal behaviors in pathophysiological processes. In addition, the spectral decomposition and wavelet analysis of intermittent forcing was used to investigate the composition of the forcing terms, which is potentially related to the long time-scale dynamics of sympathetic and parasympathetic effects regulated by our human bodies. The framework motivates the attempts to measure, understand, and control human biological systems. Further analysis can be done to investigate the consistency of intermittent behaviors and composition across patients with different pathophysiological conditions.